\documentclass[aps,tightenlines,nofootinbib,11pt]{revtex4}
\newcommand{\PRE}[1]{{#1}} % Use if preprint style
\usepackage{graphicx}
\usepackage{type1cm}
\usepackage{eso-pic}
\usepackage{color}

\usepackage{hyperref}      % hypertext links %%ARXIV
\usepackage{bm}
\usepackage{epsfig}
\def\beq{\begin{eqnarray}}
\def\eeq{\end{eqnarray}}
\def\bea{\begin{eqnarray}}
\def\eea{\end{eqnarray}}

\newcommand{\sigmath}{\sigma_{\text{th}}}

\newcommand{\OmegaDM}{\Omega_{\text{DM}}}

\newcommand{\gev}{\text{GeV}}
\newcommand{\tev}{\text{TeV}}

\newcommand{\cm}{\text{cm}}

\newcommand{\g}{\text{g}}
\newcommand{\s}{\text{s}}

\newcommand{\eg}{{\em e.g.}}

\newcommand{\eqref}[1]{Eq.~(\ref{#1})}

\newcommand{\secref}[1]{Sec.~\ref{sec:#1}}

\newcommand{\figref}[1]{Fig.~\ref{fig:#1}}

\newcommand{\Figref}[1]{Figure~\ref{fig:#1}}

\newcommand{\Omegachi}{\Omega_{\chi}}

\def\beqn{\begin{eqnarray}} 
\def\eeqn{\end{eqnarray}} 
\def\be{\begin{equation}}
\def\ee{\end{equation}}

\begin{document}

%\renewcommand{\thefootnote}{\fnsymbol{footnote}}
%\setcounter{footnote}{0}

%\preprint{number}

\title{ \PRE{\vspace*{0.15in}} {\Large Dark Matter in the Coming Decade: \\
  Complementary Paths to Discovery and Beyond} \PRE{\vspace*{0.15in}} }

%\author{CF4}
%\affiliation{
%\PRE{\vspace*{.2in}}
%}

%\author{Snowmass 2013 Cosmic Frontier Working Group 4: Dark Matter
%  Complementarity\footnote{Suggestions and corrections are most
%    welcome and should be directed to one or more of the following
%    contributors: Jim Buckley, Jonathan Feng, Manoj Kaplinghat,
%    Konstantin Matchev, Dan McKinsey, and Tim Tait.}}

\author{
    {\bf Daniel Bauer}, Fermilab;
    {\bf James Buckley}, Washington University; 
    {\bf Matthew Cahill-Rowley}, SLAC; 
    {\bf Randel Cotta}, University of California, Irvine; 
    {\bf Alex Drlica-Wagner}, SLAC;
    {\bf Jonathan L.~Feng}\footnote{Corresponding authors: {\tt  jlf@uci.edu}, {\tt  mkapling@uci.edu}, 
    {\tt matchev@phys.ufl.edu}, and {\tt ttait@uci.edu}.}, University of California, Irvine;
    {\bf Stefan Funk}, SLAC;
    {\bf JoAnne Hewett}, SLAC; 
    {\bf Dan Hooper}, Fermilab;
    {\bf Ahmed Ismail}, SLAC; 
    {\bf Manoj Kaplinghat}{${}^\ast$},  University of California, Irvine;
    {\bf Alexander Kusenko},  University of California, Los Angeles;
    {\bf Konstantin Matchev}{${}^\ast$}, University of Florida; 
    {\bf Daniel McKinsey}, Yale University;
    {\bf Tom Rizzo}, SLAC; 
    {\bf William Shepherd}, University of California, Santa Cruz;
    {\bf Tim M.~P.~Tait}{${}^\ast$}, University of California, Irvine;
    {\bf Alexander M. Wijangco}, University of California, Irvine; 
    {\bf Matthew Wood}, SLAC
    \\
on behalf of the Snowmass 2013 Cosmic Frontier Working Groups 1-4}

\date{19 February 2015}

\noaffiliation
%\email{name@place.edu}
%\affiliation{Place
%\PRE{\vspace*{.4in}}
%}

%\begin{abstract}
%\PRE{\vspace*{.3in}} Abstract.
%\end{abstract}

%\pacs{95.35.+d}
%%95.35.+d Dark matter

\maketitle

\section{Introduction}
\label{sec:intro}

Dark matter is five times as prevalent as normal matter in the 
Universe, but its identity is unknown. 
%The energy in dark matter is about five times that in normal matter in
%the visible Universe, but its identity is unknown.  
Its mere existence
implies that our inventory of the basic building blocks of nature is
incomplete, and uncertainty about its properties clouds attempts to
fully understand how the Universe evolved to its present state and how
it will evolve in the future.  Dark matter is therefore a grand
challenge for both fundamental physics and astronomy.  At the same
time, groundbreaking experiments are set to transform the field of
dark matter in the coming decade.  This prospect has drawn many new
researchers to the field, which is now characterized by an
extraordinary diversity of approaches unified by the common goal of
discovering the identity of dark matter.

As we will discuss, a compelling solution to the dark matter problem
requires synergistic progress along many lines of inquiry.  Our
primary conclusion is that the diversity of possible dark matter
candidates requires a balanced program based on four pillars: direct
detection experiments that look for dark matter interacting in the
lab, indirect detection experiments that connect lab signals to dark
matter in our own and other galaxies, collider experiments that elucidate the
particle properties of dark matter, and astrophysical probes sensitive to 
non-gravitational interactions of dark matter such 
as dark matter densities in the centers of galaxies and cooling of stars. 
%that
%determine how dark matter has shaped the evolution of large-scale
%structures in the Universe.

In this Report we summarize\footnote{Extended version of this report 
is available as \cite{Arrenberg:2013rzp}.} 
the many dark matter searches currently
being pursued in each of these four approaches.  The essential
features of broad classes of experiments are described, each with
their own strengths and weaknesses.  The goal of this Report is not to
prioritize individual experiments, but rather to highlight the
complementarity of the four general approaches that are required to
sustain a vital dark matter research program.  Complementarity also
exists on many other levels, of course; in particular, complementarity
{\em within} each approach is also important, but will be addressed by
the Snowmass Cosmic Frontier subgroups that focus on each approach.

In \secref{evidence} we briefly summarize what is known about dark
matter and some of the leading particle candidates.  In
\secref{probes}, we discuss four broad categories of search strategies
and summarize the current status of experiments in each area. We then
turn to the complementarity of these approaches in
\secref{complementarity}.  Conclusions are collected in
\secref{conclusions}.

\section{Evidence and Candidates}
\label{sec:evidence}

Dark matter was first postulated in its modern form in the 1930s to
explain the anomalously large velocities of galaxies in the Coma
cluster~\cite{Zwicky:1933gu}.  Evidence for dark matter has grown
steadily since then from data from galactic rotation
curves~\cite{Rubin:1970zz,Rubin:1980zd,1978PhDT.......195B},
weak~\cite{Refregier:2003ct} and strong~\cite{Tyson:1998vp} lensing,
hot gas in clusters~\cite{Lewis:2002mfa,Allen:2002eu}, the Bullet
Cluster~\cite{Clowe:2006eq}, Big Bang nucleosynthesis
(BBN)~\cite{Fields:2008}, 
%further constraints from large scale structure~\cite{Allen:2002eu}, 
distant supernovae~\cite{Riess:1998cb,Perlmutter:1998np}, 
the statistical distribution of galaxies~\cite{Tegmark:2003ud,Hawkins:2002sg}
and the cosmic microwave background (CMB)~\cite{Komatsu:2010fb,Ade:2013lta}.
Together, these data provide overwhelming evidence that the energy in
dark matter is roughly a quarter of the total energy in the visible
Universe and about five times the energy in normal matter.

All of this evidence for dark matter derives from its gravitational
pull on visible matter.  This does little to shed light on the
identity of dark matter, since all particles interact universally
through gravity.  To make progress, dark matter must be detected
through non-gravitational interactions. There are many possibilities.
For reviews, see, \eg,
Refs.~\cite{Bertone,Bergstrom:2009ib,Feng:2010gw}.

In the case of weakly interacting massive particles (WIMPs), dark
matter particles are produced in the hot early Universe and then
annihilate in pairs.  Those that survive to the present are known as
``thermal
relics''~\cite{Zeldovich:1965,Chiu:1966kg,Steigman:1979kw,Scherrer:1985zt}.
Such particles are generically predicted in models of physics beyond
the standard model, including models with
supersymmetry~\cite{Goldberg:1983nd,Ellis:1983ew} or extra spatial
dimensions~\cite{Servant:2002aq,Cheng:2002ej}.  Remarkably, if these
particles interact through the weak interactions of the standard
model, the resulting thermal relic density is $\Omega_X \sim {\cal
  O}(0.1)$, just right to be dark matter.  This coincidence, the
``WIMP miracle,'' provides strong motivation for dark matter with
masses from 10 GeV to 1 TeV (or 10 to 1000 times the mass of the
proton) and weak interactions with visible particles.

An alternative possibility is asymmetric dark
matter~\cite{Nussinov:1985xr,Kaplan:1991ah,Barr:1991qn,Kaplan:2009ag}.
In this case, there is a slight excess of dark particles over dark
antiparticles in the early Universe.  These annihilate until only the
slight excess of dark particles remains.  In many models, the dark
matter asymmetry is related to the normal matter--antimatter
asymmetry, and one expects the number of dark matter particles to be
similar to the number of protons.  Since dark matter contributes roughly 
five times more to the energy density of the Universe than normal matter, 
this scenario predicts dark matter particles with mass $\sim 1 - 10~\gev$.

There are several other important dark matter candidates \cite{Kusenko:2013saa}.
Axions~\cite{Peccei:1977ur,Weinberg:1977ma,Wilczek:1977pj,Asztalos:2006kz}
are strongly motivated by a severe problem of the standard model: the
theory of the strong interactions naturally predicts large CP
violating effects that have not been observed.  Axions would resolve
this problem elegantly by suppressing CP violation to experimentally
allowed levels.  Right-handed or sterile neutrinos are motivated by the observation of 
non-zero neutrino masses, and for certain ranges of masses and interaction strengths,
they may be dark matter~\cite{Dodelson:1993je,Kusenko:2009up,Abazajian:2012ys}.
Alternatively, dark matter may be in a so-called hidden sector, which
has its own set of matter particles and forces, through which the dark
matter interacts with other currently unknown particles. 

\begin{figure}[tb]
\includegraphics[width=0.95\columnwidth]{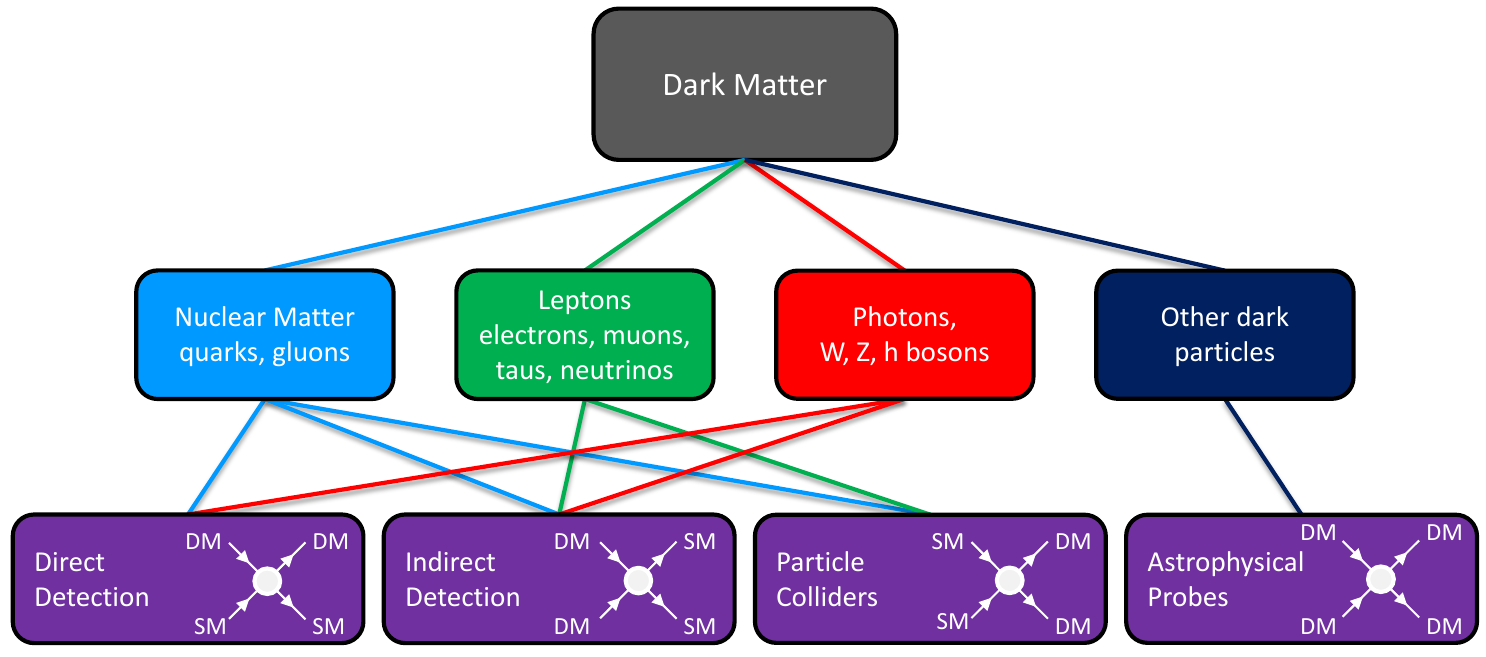}
%\vspace*{-.1in}
\caption{Dark matter may have non-gravitational interactions with one
  or more of four categories of particles: nuclear matter, leptons,
  photons and other bosons, and other dark particles.  These
  interactions may then be probed by four complementary approaches:
  direct detection, indirect detection, particle colliders, and
  astrophysical probes. The lines connect the experimental approaches
  with the categories of particles that they most stringently probe 
  (additional lines can be drawn in specific model scenarios).
  The diagrams give example reactions of dark matter (DM) with
  standard model particles (SM) for each experimental approach.
\label{fig:interactions}}
\end{figure}

Although these dark matter candidates differ in important ways, in
most cases, they have non-gravitational interactions through which they
may be detected.
%\footnote{A notable exception is primordial black holes~\cite{Carr:2003bj,Frampton:2010sw}.}  
The non-gravitational interactions may be with any
of the known particles or, as noted above for hidden sector dark
matter, with other currently unknown particles.  These possibilities
are shown in \figref{interactions}, where the particles are grouped
into four categories: nuclear matter; leptons; photons and other
bosons; and other as-yet unknown particles.  Dark matter may interact
with one type of particle, or it may interact with several. 

A complete research program in dark matter therefore requires a
diverse set of experiments that together probe all possible types of
couplings.  At present, the experiments may be grouped into the
following four categories:
\begin{itemize}
\setlength{\itemsep}{1pt}\setlength{\parskip}{0pt}\setlength{\parsep}{0pt}
\item {\em Direct Detection}.  Dark matter scatters off a detector,
  producing a detectable signal.  Prime examples are the detection of
  WIMPs through scattering off nuclei and the detection of axions
  through their interaction with photons in a magnetic field.
\item {\em Indirect Detection}.  Pairs of dark matter particles
  annihilate producing high-energy particles (antimatter, neutrinos,
  or photons).  Alternatively, dark matter may be metastable, and its
  decay may produce the same high-energy particles.
\item {\em Particle Colliders}.  Particle colliders, such as the Large
  Hadron Collider (LHC) and proposed future lepton colliders, produce
  dark matter particles, which escape the detector, but are
  discovered as an excess of events with missing energy or momentum.
\item {\em Astrophysical Probes}.  The particle properties of dark
  matter are constrained through its impact on astrophysical
  observables.  Examples include self-interaction of dark matter particles 
affecting central dark matter densities in galaxies (inferred from rotation 
velocity or velocity dispersion measures), mass of dark matter particle affecting 
dark matter substructure in galaxies (inferred from strong lensing data) and 
annihilation of dark matter in the early Universe affecting the Cosmic 
Microwave Background fluctuations.
\end{itemize}
These search strategies are shown in \figref{interactions} and are
connected to the particle interactions they most stringently probe.
In the next Section, we briefly describe these four approaches and
summarize their current status.

\section{The Four Pillars of Dark Matter Detection}
\label{sec:probes}

\subsection{Direct Detection}
\label{sec:direct}

Dark matter permeates the whole Universe, and its local density on
Earth is known to be $5\times 10^{-25}~\g/\cm^3$ to within a factor of 2.
This creates the opportunity to detect dark matter particles {\em
  directly} as they pass through and scatter off normal
matter~\cite{Goodman:1984dc}.  Such events are extremely rare, and so
the direct detection approach requires sensitive detectors with
exquisite background rejection \cite{Cushman:2013zza}.  The expected signals depend on the
nature of the dark matter particles and their interactions.  For a
list of current and planned experiments, see~\cite{TableDD}.

In the case of WIMPs, direct searches are extremely promising.
Experimental techniques include detectors that record ionization,
scintillation light, and phonons.  The most sensitive of the detectors
employ multiple techniques, and the interplay of each is used to
discriminate against backgrounds.  Depending on the target material,
experiments can be sensitive to (a combination of) spin-dependent and
spin-independent WIMP interactions with matter. The sensitivity of the
current generation of detectors for spin-independent (spin-dependent) cross sections
for scattering off protons is approaching 
$\sigma_{\text{SI}}^p \sim 10^{-45}~\cm^2$ 
($\sigma_{\text{SD}}^p \sim 10^{-39}~\cm^2$)
for WIMP masses of $\sim 100~\gev$, with orders of
magnitude improvement expected in the coming decade \cite{Cushman:2013zza}.  
In the last years a number of experiments have reported potential signals that could 
be interpreted as very light ($~10$ GeV) DM particles. Although this interpretation is 
still inconclusive, it greatly motivates designing experiments with low threshold energies.
%For asymmetric
%dark matter with masses $\sim \gev$, the reduced recoil energy is
%challenging to detect, but there has been significant progress in

Axions also have strong prospects for direct detection. Cosmological
and astrophysical constraints restrict the allowed axion mass range to
be between 1 $\mu$eV and 1 meV.  In a static magnetic field, there is
a small probability for cosmologically produced axions to be converted
by virtual photons to real microwave photons by the Primakoff
effect~\cite{Sikivie:1983ip}. This would produce a monochromatic
signal with a line width of $dE/E\sim 10^{-6}$, which could be detected
in a high-$Q$ microwave cavity tunable over GHz frequencies.  
In the near future, these searches will be sensitive to models with 
axion mass $\sim \mu$eV, which is the favored mass range if axions 
are a significant component of dark matter.

\subsection{Indirect Detection}
\label{sec:indirect}

In contrast to direct detection experiments, indirect detection
efforts do not aim to detect dark matter particles
themselves. Instead, they attempt to detect the standard model
particles that are produced in their annihilations or decays. Signals
for indirect detection experiments include photons (gamma rays,
X-rays, radio), neutrinos and cosmic rays (including positrons,
electrons, antiprotons, and antideuterons). Many types of detectors
and telescopes have been designed and deployed with these goals in
mind, ranging from space- and ground-based gamma-ray telescopes and
cosmic-ray detectors, to large underground, under-ice, and underwater
neutrino telescopes \cite{Buckley:2013bha}.  Current and planned indirect search experiments
are listed in~\cite{TableID}.

Motivating the existing and planned indirect detection efforts is the
characteristic annihilation cross section of WIMP thermal
relics. Although the precise value of this cross section depends on a
number of model-dependent features, WIMP candidates that annihilate to
the correct relic density to be dark matter typically have average
cross sections (multiplied by the relative velocity of the
annihilating WIMPs) of $\langle \sigma v \rangle\sim 3 \times
10^{-26}~\cm^3/\s$.

Excitingly, indirect detection experiments have started to reach the
level of sensitivity required to discover WIMPs with this annihilation
cross section (thermal WIMP). Current constraints from the Fermi
Gamma-Ray Space Telescope observations of some Milky Way satellite
galaxies have begun to exclude the low-mass region of some thermal WIMP
models~\cite{GeringerSameth:2011iw}. 
On the other hand, there are excesses seen in gamma rays and microwaves 
towards the Galactic Center that are remarkably consistent with a 
thermal WIMP dark matter interpretation, though other explanations have not been ruled out.

Measurements of the cosmic-ray antiproton spectrum from PAMELA 
also constrain some of the thermal WIMP
models, with further results expected from AMS-02 experiment on the
International Space Station.  Above about 10 GeV, there is an excess 
in the cosmic-ray positron flux which is seen by PAMELA, Fermi and 
the AMS-02 experiments. This has been interpreted as a sign of dark matter
annihilating to leptons, although nearby pulsars could also provide an explanation.  
The indirect detection of dark matter is a major science goal of the
kilometer-scale neutrino telescope IceCube as well. In contrast
to other indirect searches, neutrino telescopes are most sensitive to
WIMPs that annihilate in the core of the Sun. Current constraints from
IceCube data have begun to exclude otherwise viable WIMP models.

Indirect searches are not limited to dark matter in the form of
WIMPs. Sterile neutrinos, for example, are predicted to decay, leading
to potentially observable X-ray spectral lines~\cite{Abazajian:2001vt}. 
Solar axion searches are sensitive to axions produced in the Sun 
(because of their coupling to photons) which then convert in the 
magnetic field of the detector \cite{Kusenko:2006rh}.

\subsection{Particle Colliders}
\label{sec:colliders}

Dark matter may also be produced in high-energy particle collisions.
For example, if dark matter has substantial couplings to nuclear
matter, it can be created in proton-proton collisions at the
Large Hadron Collider (LHC).  Once produced, dark matter particles
will likely pass through detectors without a trace, but their
existence may be inferred from an imbalance in the visible momentum,
just as in the case of neutrinos.  Searches for dark matter at the LHC
are therefore typified by missing momentum, and can be categorized by
the nature of the visible particles that accompany the dark matter
production.  Because backgrounds are typically smaller for larger
values of missing momentum, collider searches tend to be most
effective for low-mass dark matter particles, which are more easily
produced with high momentum.

There are two primary mechanisms by which the LHC could hope to
produce dark matter together with hadronic jets (see, e.g., Chapters
13 and 14 in Ref.~\cite{Bertone}).  In the first, two
strongly interacting parent particles of the dark matter theory are
produced, and each one subsequently decays into the dark matter and
standard model particles, resulting in missing momentum plus two or
more jets of hadrons.  Since the production relies on the strong
force, the rate of production is specified by the color charge, mass,
and spin of the parent particles and is rather insensitive
to the mass of the dark matter itself.  Current null results from LHC
searches for the supersymmetric partners of quarks typically exclude 
such particles with masses less than $\sim 1.5~\tev$, although this
limit is weakened in the case of mass-degenerate spectra.

A second mechanism produces the dark matter directly
together with additional radiation from the initial quarks or gluons
participating in the reaction, resulting in missing momentum recoiling
against a single ``mono-jet.''  Since this process does not rely as
explicitly on the existence of additional colored particles that
decay into dark matter, it is somewhat less sensitive to the details
of the specific theory and places bounds directly in the parameter
space of the dark matter mass and interaction strength.  However, one
does need to posit a specific form of the interaction between the dark
matter and quarks or gluons.  For electroweak-size couplings and
specific choices of the interaction structure, these searches exclude
dark matter masses below about 500 GeV.

High energy lepton colliders may create dark matter through analogous
processes, such as production of dark matter along with a photon
radiated from the initial leptons.  For electroweak-size couplings of
dark matter to electrons, LEP excluded dark matter masses below about
90 GeV.  A future high-energy lepton collider could conceivably
discover dark matter particles with masses up to roughly half the
collision energy, \eg, 500 GeV for a 1 TeV ILC.  For a list of current
and proposed future colliders, see~\cite{TableEF}.

\subsection{Astrophysical Probes}
\label{sec:astrophysics}

Non-gravitational interactions of dark matter affect a variety of
astrophysical observables, including the number density and internal
structure of galaxies. The majority of dark matter searches are
focused on candidates that are astrophysically categorized as cold and
collisionless dark matter (CDM).  CDM predictions agree amazingly well
with cosmological data~\cite{Komatsu:2010fb}, but the observed
densities of dark matter in the central parts of galaxies are often
lower than the simplest predictions.

Evidence for lowered central densities and constant central density
cores is seen in the least massive to the most massive
self-gravitating objects in the Universe, including satellite galaxies
of the Milky Way, spiral galaxies, and clusters of galaxies. Constant
density cores of dark matter are in conflict with the simplest CDM
predictions, but baryonic feedback (such as outflows from supernovae)
may change those predictions. Both feedback and deviations from CDM
paradigm such as warm dark matter (WDM) or strongly self-interacting
dark matter (SIDM) may be required to explain the data fully.

Compared to CDM, models of WDM have reduced power in small-scale 
density fluctuations and the number of low-mass dark matter halos is
dramatically reduced.
Hidden sector models where dark matter particles interact with other light 
particles can also lead to the same effects. 
These predictions can be tested in the future using strong
gravitational lensing systems, precise observations of the clustering in 
the Universe and searches for new satellite galaxies and stellar streams.  
The mass-scale below which halo formation is suppressed is directly related to 
%the ``warmth'' (primordial temperature),  which in turn is related to 
one or more parameters of the particle physics model --- for example, 
mass and couplings of sterile neutrino dark matter 
candidates~\cite{Kusenko:2006rh}.  
In addition, the central density of dark matter halos is also reduced in WDM 
cosmology, but large constant density cores have not been shown to form.

In comparison, the primary effect of SIDM is to reduce the central
density of dark matter halos and create constant density (spherical)
cores with observable effects if the cross section to particle mass
ratio is in the $0.1-1~\cm^2/\g$ range~\cite{Spergel:1999mh}. 
This range is consistent with measured dark matter halo shapes but larger
cross sections are excluded by Bullet Cluster
observations~\cite{Clowe:2006eq}, and the predictions for this range 
may be tested with more sensitive observations of the Milky Way satellites 
at the faint end and clusters of galaxies at the bright end. Cross sections of this 
magnitude can be produced in hidden sector dark matter models through the
exchange of a light gauge boson and this interaction can also endow
the dark matter particle with the right relic density through a hidden
sector analogue of the WIMP miracle~\cite{Feng:2009hw}. 

In addition to structure formation, non-gravitational interactions of
dark matter could impact a variety of other astrophysical
phenomena. Coupling of axions and light sterile neutrinos (or
generally any light hidden-sector particles) to standard model
particles may affect the cooling of compact objects (stars, neutron
stars, white dwarfs, supernovae) or the transparency of extragalactic
background light to high-energy photons, which leads to stringent
constraints on models~\cite{Raffelt:2000kp}.  The observed fluctuations
in the Cosmic Microwave Background (WMAP, Planck) are sensitive enough
to the (small) dark matter annihilation rate during the era of recombination
(age $\sim 380,000~{\rm years}$) to be able to constrain thermal WIMP
models~\cite{Padmanabhan:2005es}. In the same vein, decays of dark matter particles (with 
lifetime much larger than the age of the Universe) can affect the reionization of neutral atoms  
about 13 billion years ago. 

While dark matter physics may have imprinted tell-tale astrophysical
signatures, it will be hard to unambiguously identify such signatures
as non-gravitational interactions of dark matter. The complementarity
with direct, indirect or collider searches is an essential part of
this endeavor.

\section{Complementarity}
\label{sec:complementarity}

\subsection{Qualitative Complementarity}
\label{sec:basic}

As evident from the brief descriptions in \secref{probes}, every
experimental approach provides useful information for every dark
matter scenario.  At the same time, each approach is subject to
different systematic uncertainties and no approach will illuminate all
aspects of dark matter.  In detail, what is learned from each approach
is highly scenario-dependent.

At a qualitative level, the complementarity may be illustrated by the
following observations that follow from basic features of each
approach:
\begin{itemize}
\setlength{\itemsep}{1pt}\setlength{\parskip}{0pt}\setlength{\parsep}{0pt}
\item {\em Direct Detection} is perhaps the most straightforward
  detection method, with excellent prospects for improved sensitivity
  in the coming decade and for discovering WIMPs.  The approach requires
  careful control of low-energy backgrounds, and is relatively
  insensitive to dark matter that couples to leptons only, or to
  WIMP-like dark matter with mass $\sim 1~\gev$ or below.
\item {\em Indirect Detection} is sensitive to dark matter
  interactions with all standard model particles, directly probes the
  annihilation process suggested by the WIMP miracle, and experimental
  sensitivities are expected to improve greatly on several fronts in
  the coming decade.  Discovery through indirect detection requires
  understanding astrophysical backgrounds and the signal strength is
  typically subject to uncertainties in halo profiles. Indirect
  detection signals are absent if dark matter annihilation is
  insignificant now, for example, as in the case of asymmetric dark
  matter.
\item {\em Particle Colliders} provide the opportunity to study dark
  matter in a highly controlled laboratory environment, may be used to
  precisely constrain many dark matter particle properties, and are
  sensitive to the broad range of masses favored for WIMPs.  Hadron
  colliders are relatively insensitive to dark matter that interacts
  only with leptons, and colliders are unable to distinguish missing
  momentum signals produced by a particle with lifetime $\sim 100$ ns
  from one with lifetime $\agt 10^{17}~\s$, as required for dark
  matter.
\item {\em Astrophysical Probes} are unique probes of the ``warmth''
  of dark matter and hidden dark matter properties, such as its
  self-interaction strength, and they measure the effects of
  dark matter properties on structure formation in the Universe.
  Astrophysical probes are typically unable to distinguish various
  forms of CDM from each other or make other precision measurements of
  the particle properties of dark matter.
\end{itemize}

\subsection{Quantitative Complementarity}
\label{sec:quantitative}

\subsubsection{Effective Operator Description}

The qualitative features outlined above may be illustrated in a simple
and fairly model-independent setting by considering dark matter that
interacts with standard model particles through four-particle contact
interactions, which represent the exchange of very heavy particles.
These contact interactions are expected to work well to describe
theories in which the exchanged particle mass is considerably
larger than the momentum transfer of the physical process of interest.

To do this, we may choose representative, generation-independent, couplings of a spin-1/2 dark
matter particle $\chi$ with quarks $q$, gluons $g$, and leptons $\ell$ (including neutrinos)
given by
\begin{equation}
\frac{1}{M_q^2} ~\bar{\chi} \gamma^\mu\gamma_5 \chi \sum_q \bar{q} \gamma_\mu\gamma_5  q
+ \frac{\alpha_S}{M_g^3}~ \bar{\chi}  \chi G^{a \mu \nu} G^a_{\mu \nu}
+ \frac{1}{M_\ell^2} ~
\bar{\chi} \gamma^\mu  \chi \sum_\ell \bar{\ell} \gamma_\mu  \ell~.
\label{lagrangian}
\end{equation}
The interactions with quarks mediate spin-dependent direct signals,
whereas those with gluons mediate spin-independent direct signals.
The coefficients $M_q$, $M_g$, and $M_\ell$ characterize the strength
of the interaction with the respective standard model particle, and in
this representative example should be chosen such that the combined
annihilation cross section into all three channels provides the
correct relic density of dark matter.  The values of the three
interaction strengths, together with the mass of the dark matter
particle $m_\chi$, completely define this theory and allow one to
predict the rate of both spin-dependent and spin-independent direct
scattering, the annihilation cross section into quarks, gluons, and
leptons, and the production rate of dark matter at colliders.

\begin{figure}[tb]
\includegraphics[width=0.32\columnwidth]{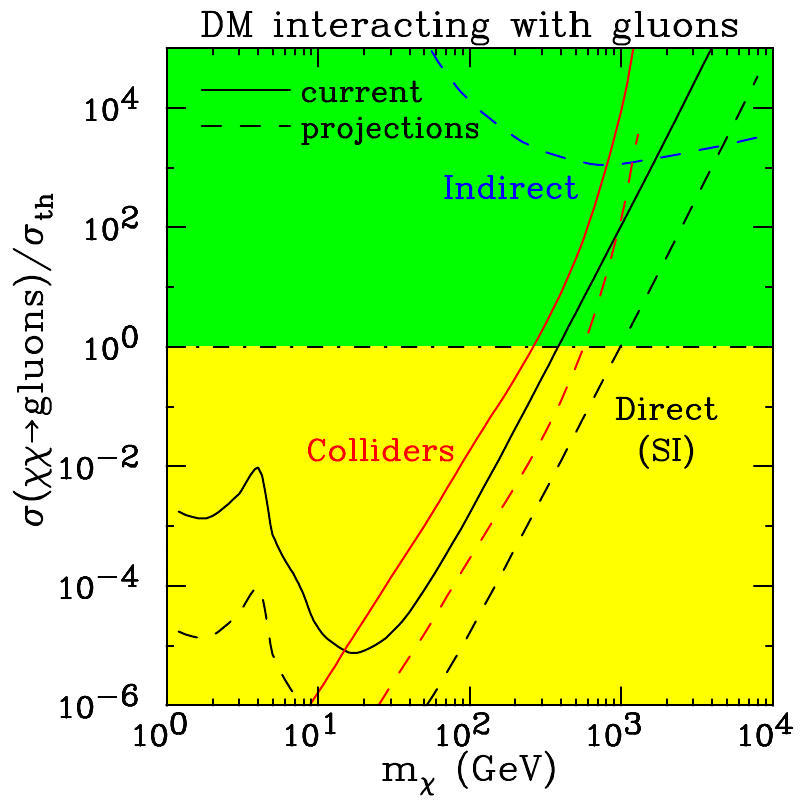}
\includegraphics[width=0.32\columnwidth]{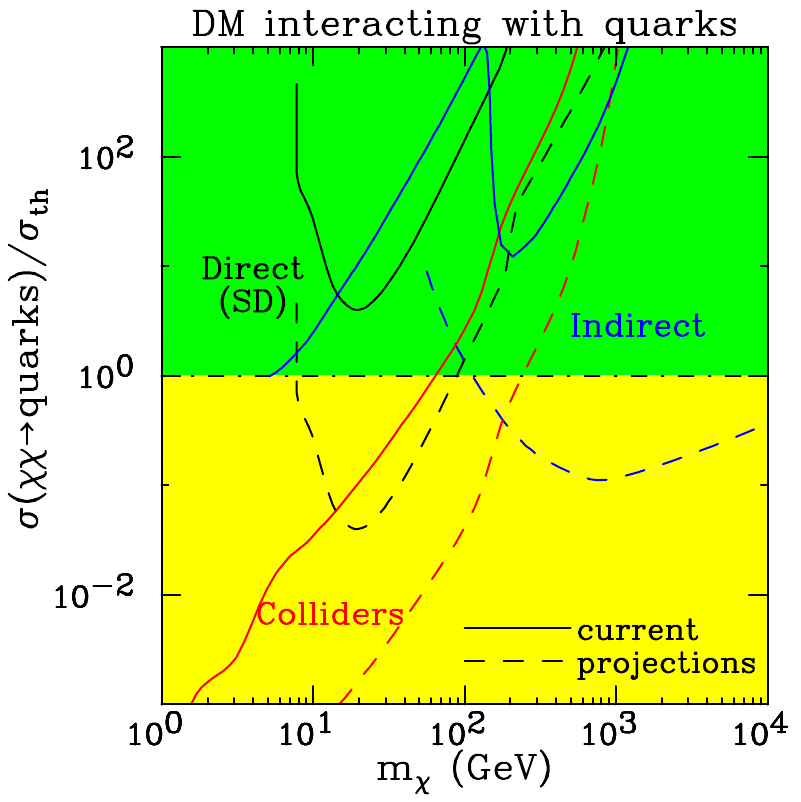}
\includegraphics[width=0.32\columnwidth]{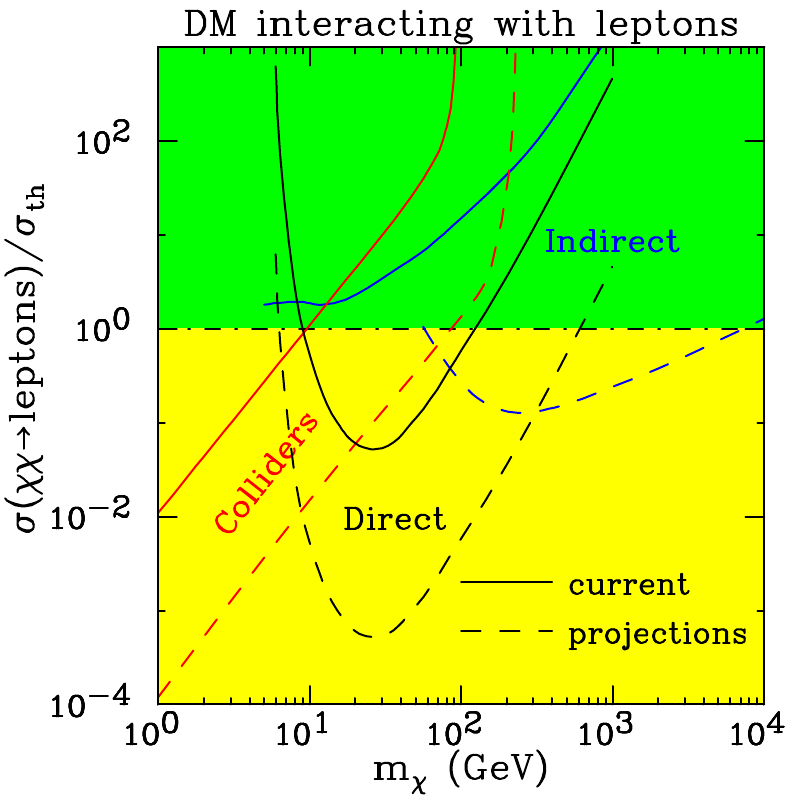}
%\hfil
%\includegraphics[width=0.49\columnwidth]{DMQuarkSDOps_sigma} \\
%\includegraphics[width=0.49\columnwidth]{DMLeptonicOps_sigma} 
%\vspace*{-.1in}
\caption{Dark matter discovery prospects in the $(m_{\chi},
  \sigma/\sigmath)$ plane for current and future direct
  detection~\cite{DMtools}, indirect
  detection~\cite{Ackermann:2011wa,IndirectCTA}, and particle
  colliders ~\cite{Chatrchyan:2012me,ATLAS:2012ky,Vacavant:2001sd} for
  dark matter coupling to gluons~\cite{Goodman:2010ku},
  quarks~\cite{Goodman:2010ku,Beltran:2008xg}, and
  leptons~\cite{Fox:2011fx,Chae:2012bq}, as indicated.
\label{fig:prospects}}
\end{figure}

Each class of dark matter search outlined in \secref{probes} is
sensitive to some range of the interaction strengths for a given dark
matter mass.  Therefore, they are all implicitly putting a bound on
the annihilation cross section into a particular channel.  Since the
annihilation cross section predicts the dark matter relic density, the
reach of any experiment is thus equivalent to a fraction of the
observed dark matter density.  This connection can be seen in the
plots in \figref{prospects}, which show the annihilation cross section
normalized to the value $\sigmath$, which is required\footnote{For 
non-thermal WIMPs, e.g. asymmetric DM, the annihilation cross-section
does not have a naturally preferred value, but the plots in \figref{prospects}
are still meaningful.} for a thermal
WIMP to account for all of the dark matter in the Universe. If the
discovery potential for an experiment with respect to one of the
interaction types reaches cross sections below $\sigmath$ (the
horizontal dot-dashed lines in \figref{prospects}), that experiment will
be able to discover thermal relic dark matter that interacts only with
that standard model particle and nothing else.

If an experiment were to observe an interaction consistent with an
annihilation cross section below $\sigmath$ (yellow-shaded regions in
\figref{prospects}), it would have discovered dark matter but we would
infer that the corresponding relic density is too large, and therefore
there are important annihilation channels 
(or new physics that changes early Universe cosmology) still waiting to be discovered.  
Finally, if an experiment were to observe a cross section
above $\sigmath$ (green-shaded regions in \figref{prospects}), it
would have discovered one species of dark matter, which, however,
could not account for all of the dark matter (within this model
framework), and consequently point to other dark matter species still
waiting to be discovered.

In \figref{prospects}, we assemble the discovery potential and current
bounds for several near-term dark matter searches that are sensitive
to interactions with quarks and gluons, or leptons.  It is clear that
the searches are complementary to each other in terms of being
sensitive to interactions with different standard model particles.
These results also illustrate that within a given interaction type,
the reach of different search strategies depends sensitively on the
dark matter mass.  For example, direct searches for dark matter are
very powerful for masses around 100 GeV, but have difficulty at very
low masses, where the dark matter particles carry too little momentum
to noticeably affect heavy nuclei.  This region of low mass is
precisely where collider production of dark matter is easiest, since
high energy collisions readily produce light dark matter particles 
with large momenta.

\subsubsection{Complete Models}

The effective theory description (\ref{lagrangian}) of the dark matter
interactions with standard model particles is an attempt to capture
the salient features of the dark matter phenomenology without
reference to any specific theoretical model.  However, the
complementarity between the different dark matter probes seen in
\figref{prospects} persists also when one considers specific
well-motivated theoretical models.  Among the many possible alternatives,
low energy supersymmetry~\cite{Martin:1997ns} has been the most
popular and widely studied extension of the standard model, and we
shall use it here as our second example. In supersymmetry, the
DM candidate is generally the lightest neutralino $\tilde\chi^0_1$,
which is its own anti-particle.

\begin{figure}[t]
\includegraphics[width=0.45\columnwidth]{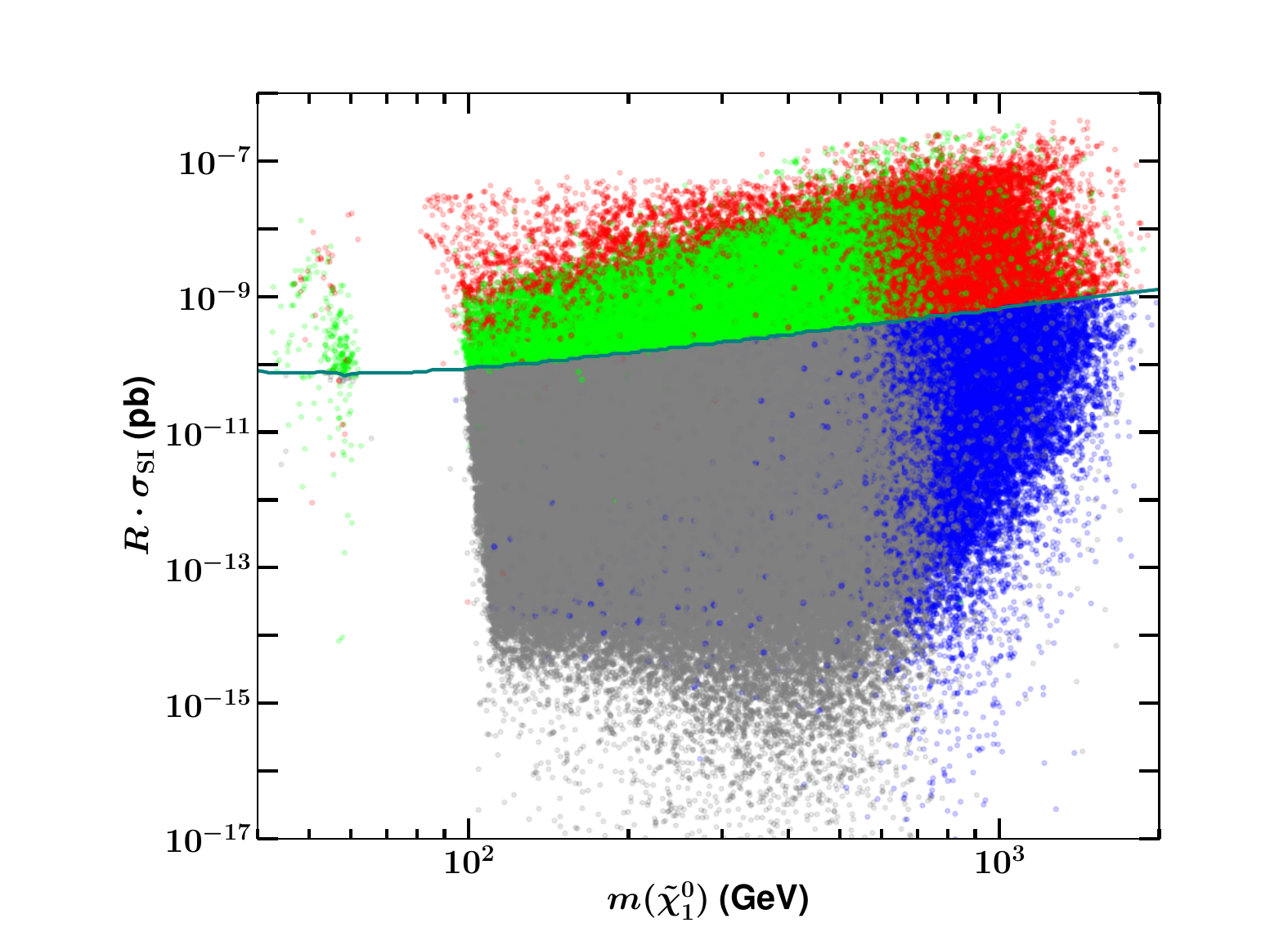}
\includegraphics[width=0.45\columnwidth]{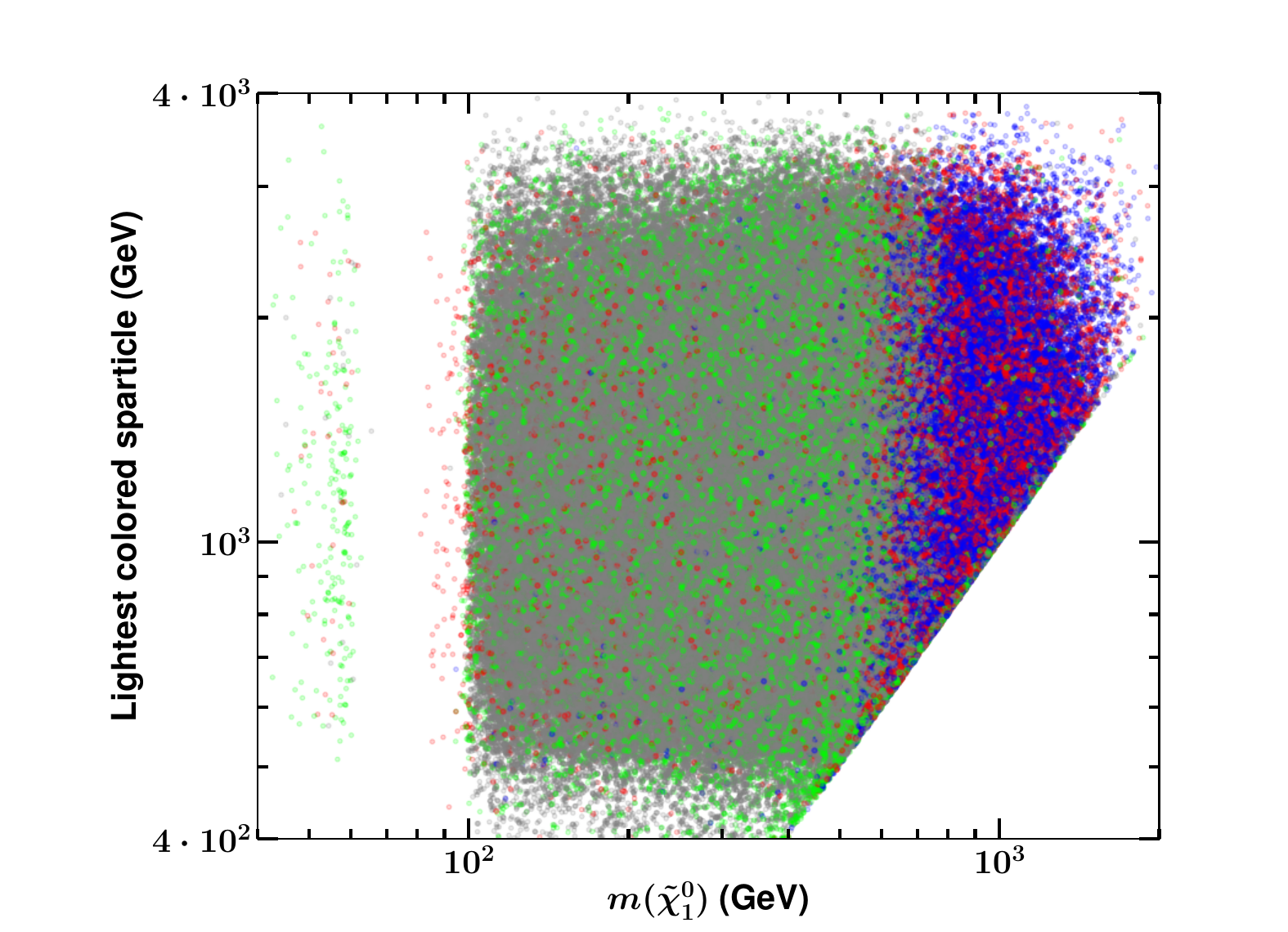}
\caption{Results from a model-independent
  scan~\cite{CahillRowley:2012cb,pMSSMwhitepaper}
%,CahillRowley:2012rv,CahillRowley:2012kx} 
of the full parameter space in the minimal supersymmetric model
(MSSM), presented in the $(m_\chi,R\cdot\sigma_{\text{SI}}^p)$ plane (left panel) 
or the $(m_\chi,m_{\text{LCSP}})$ plane (right panel), where $R=\Omegachi / \OmegaDM$. 
The models are divided into categories, depending on whether
dark matter can be discovered in future direct detection experiments
(green points), indirect detection experiments (blue points) or both
(red points). The gray points represent models that may be discovered at 
the upgraded LHC, but escape detection in future direct or indirect detection 
experiments. 
\label{fig:pMSSM}}
\end{figure}

Even within the general framework of supersymmetry, there are many
different model scenarios, distinguished by a number of input
parameters ($\sim 20$). A model-independent approach to supersymmetry
is to scan over all those input parameters and consider all models
that pass all existing experimental constraints and have a dark matter
candidate which could explain at least a portion of the observed dark
matter density~\cite{CahillRowley:2012cb}.
%,CahillRowley:2012rv,CahillRowley:2012kx}.  
Results from such model-independent scans with over 200,000 points are
shown in \figref{pMSSM}, where each dot represents one particular
supersymmetric model. Within each model, the dark matter interactions
are completely specified and one can readily compute all relevant dark
matter signals. The models are categorized depending on the
observability of a dark matter signal in direct detection
experiments (green points), indirect detection experiments (blue points)
or both (red points). The gray points represent models that escape
detection in dark matter experiments, but may be discovered at the
upgraded LHC, if the mass of the lightest colored superpartner is
within $m_{\text{LCSP}} \sim 3$ TeV. A sizable fraction of models
(the blue points) can only be seen in indirect
detection (via ground-based gamma ray telescopes).  Another large
fraction of models (the gray points) can only be
seen at the LHC. \Figref{pMSSM} demonstrates that the three different
dark matter probes nicely combine to discover many (albeit not all)
supersymmetry models in this scan.
 
\subsection{Post-Discovery Complementarity}
\label{sec:postdiscovery}

As important as a broad program of complementary searches is to
establishing a compelling signal for dark matter, it becomes even more
important after a signal has been reported for several reasons.

First, as is well known, many tentative dark matter signals have
already been reported.  The potential identification of a quarter of
the Universe will require extraordinary proof in the form of
verification by other experiments.

Second, each search strategy has its limitations.  For example, as
noted in \secref{basic}, the discovery of a dark matter signal at
particle colliders only establishes the production of a particle with
lifetime greater than about 100 ns.  The assumption that this particle
contributes to dark matter requires an extrapolation in lifetime of 24
orders of magnitude! It is only by corroborating a particle collider
discovery through another method that one can claim that the collider
discovery is relevant for cosmology.

Last, the discovery of dark matter will usher in a rich and
decades-long program of dark matter studies. Consider the following
(somewhat optimistic) scenario: The LHC sees a missing energy signal, and precision
measurements find evidence that it is due to a 60 GeV neutralino.
This result is confirmed by direct search experiments, which discover
a signal consistent with this mass.  However, further LHC and ILC
studies constrain the neutralino's predicted thermal relic density $\Omegachi$ to
be half of $\Omega_{\text{DM}}$, implying that it is not a thermal
relic, or that it makes up only half of the dark matter.  The puzzle
is resolved when axion detectors discover a signal, which is
consistent with axions making up the rest of the dark matter, and
progress in astrophysical theory, simulations, and observations leads to
a consistent picture with dark matter composed entirely of CDM.  The combined
data establish a new standard cosmology in which dark matter is
composed of equal parts neutralinos and axions, and extend our
understanding of the early Universe back to neutralino freezeout, just
1 ns after the Big Bang.  Direct and indirect detection rates are then
used to constrain the local dark matter density, halo profiles, and
substructure, establishing the new fields of neutralino and axion
astronomy.

This two-component scenario is more complicated than assumed in many
dark matter studies, but it is still relatively simple --- as is often
noted, the visible Universe has many components, and there is no
reason that the dark Universe should be any simpler.  As simple as
this scenario is, however, it illustrates the point that, even for
dark matter candidates that we have studied and understood, the
information provided by several approaches will be essential to
understanding the particle nature of dark matter and its role in
astrophysics and cosmology. A balanced program with components in each
of the four approaches is required to cover the many well-motivated
dark matter possibilities, and their interplay will likely be
essential to realize the full potential of upcoming discoveries.

\section{Conclusions}
\label{sec:conclusions}

The problem of identifying dark matter is central to the fields of
particle physics and astrophysics, and has become a leading problem in
all of basic science.  In the coming decade, the field of dark matter
will be transformed, with a perfect storm of experimental and
technological progress set to put the most promising ideas to the
test.

Dark matter searches rely on four approaches or pillars: direct
detection, indirect detection, particle colliders, and astrophysical
probes.  In this Report, we have described the complementary relation
of these approaches to each other.  This complementarity may be seen
on several levels.  First, these approaches are qualitatively
complementary: they differ in essential characteristics, and they rely
on different dark matter properties to see a signal.  A complementary
set of approaches is required to be sensitive to the dark matter
possibilities that are currently both viable and well-motivated.  The
approaches are also quantitatively complementary: within a given class
of dark matter possibilities, these approaches are sensitive to
different dark matter interactions and mass ranges.

Last, the discovery of a compelling dark matter signal is only the
beginning.  Complementary experiments are required to verify the
initial discovery, to determine whether the particle makes up all of
dark matter or only a portion, and to identify its essential
properties, such as its interactions, spin, and mass, and to determine
its role in forming the large scale structures of the Universe that we
see today.  A balanced dark matter program is required to carry out
this research program to discover and study dark matter and to
transform our understanding of the Universe on both the smallest and
largest length scales.

%\newpage

\bibliography{bibdmcomp}{}

\end{document}